\documentclass[preprintnumbers,amsmath,amssymb,onecolumn,superscriptaddress]{revtex4-1}
\usepackage{graphicx}
\usepackage{dcolumn}
\usepackage{bm}
\usepackage{amssymb}
\usepackage{amsmath}
\usepackage{epstopdf}
\usepackage{booktabs}
\usepackage{threeparttable}
\usepackage[pdfstartview=FitH, CJKbookmarks=true, bookmarksnumbered=true, bookmarksopen=true, colorlinks, pdfborder=001, linkcolor=blue, anchorcolor=blue, citecolor=blue]{hyperref}

\begin{document}

\title{Evidence for nodal superconductivity in a layered compound Ta$_4$Pd$_3$Te$_{16}$}

\author{G. M. Pang}
\affiliation{Center for Correlated Matter and Department of Physics, Zhejiang University, Hangzhou 310058, China}
\author{M. Smidman}
\email{msmidman@zju.edu.cn}
\affiliation{Center for Correlated Matter and Department of Physics, Zhejiang University, Hangzhou 310058, China}
\author{W. H. Jiao}
\affiliation{Department of Physics, Zhejiang University of Science and Technology, Hangzhou 310023, China}
\author{L. Jiao}
\affiliation{Center for Correlated Matter and Department of Physics, Zhejiang University, Hangzhou 310058, China}
\author{Z. F. Weng}
\affiliation{Center for Correlated Matter and Department of Physics, Zhejiang University, Hangzhou 310058, China}
\author{W. B. Jiang}
\affiliation{Center for Correlated Matter and Department of Physics, Zhejiang University, Hangzhou 310058, China}
\author{C. Y. Guo}
\affiliation{Center for Correlated Matter and Department of Physics, Zhejiang University, Hangzhou 310058, China}
\author{Y. Chen}
\affiliation{Center for Correlated Matter and Department of Physics, Zhejiang University, Hangzhou 310058, China}
\author{G. H. Cao}
\affiliation{Center for Correlated Matter and Department of Physics, Zhejiang University, Hangzhou 310058, China}
\affiliation{Collaborative Innovation Center of Advanced Microstructures, Nanjing 210093, China}
\author{H. Q. Yuan}
\email{hqyuan@zju.edu.cn}
\affiliation{Center for Correlated Matter and Department of Physics, Zhejiang University, Hangzhou 310058, China}
\affiliation{Collaborative Innovation Center of Advanced Microstructures, Nanjing 210093, China}

\date{\today}

\begin{abstract}
We report an investigation of the London penetration depth $\Delta\lambda(T)$ on single crystals of the layered superconductor Ta$_4$Pd$_3$Te$_{16}$, where the crystal structure has quasi-one-dimensional characteristics. A linear temperature dependence of $\Delta\lambda(T)$ is observed for $T\ll T_c$, in contrast to the exponential decay of fully gapped superconductors. This indicates the existence of line nodes in the superconducting energy gap. A detailed analysis shows that the normalized superfluid density $\rho_s(T)$, which is converted from $\Delta\lambda(T)$, can be  well described by a multigap scenario, with nodes in one of the superconducting gaps, providing clear evidence for nodal superconductivity in Ta$_4$Pd$_3$Te$_{16}$.

\end{abstract}

\maketitle

\section{Introduction}

Recently the layered Pd-based ternary chalcogenides have attracted much research interest. New compounds with similar crystal structures were discovered, which opened up new opportunities to investigate the relationship between superconductivity and reduced crystal dimensionality \cite{zhang2013anomalous,yu2013superconducting,jiao2016superconductivity}. Unusual superconducting properties were revealed in Ta$_2$Pd$_x$S$_5$ ($x \lesssim$ 1.0) and Nb$_2$Pd$_{0.81}$S$_5$, where extremely large upper critical fields of $\mu_0$H$_{c2}$(0)~=~31~T and $\mu_0$H$_{c2}$(0)~=~37~T were observed respectively, both of which are almost twice the size of  the Pauli limiting field ($H_P$) \cite{lu2013superconductivity,zhang2013superconductivity}. These are reminiscent of the quasi-one-dimensional (Q1D) organic compounds (TMTSF)$_2X$ (TMTSF = tetramethyltetraselenafulvalene, $X$~=~PF$_6$, ClO$_4$), which are believed to be unconventional superconductors \cite{lee1997anisotropy,oh2004magnetic}. Moreover,  subsequent specific heat measurements in Nb$_2$Pd(S$_{1-x}$Se$_x$) and Ta$_2$PdSe$_5$ show a slight deviation from the typical behavior of single band $s$-wave superconductors, which likely indicates multi-band superconductivity in these systems \cite{niu2013effect,zhang2015superconductivity}.

A  new layered Pd-based ternary compound Ta$_4$Pd$_3$Te$_{16}$ was found to exhibit superconductivity at $T_c\approx$4.6~K \cite{jiao2014superconductivity}. It possesses a layered crystal structure as well as quasi-one-dimensional (Q1D) characteristics, with chains running along the $b$ axis. Band structure calculations for Ta$_4$Pd$_3$Te$_{16}$ reveal that its Fermi surface consists of four branches, including two one-dimensional nested sheets, a two-dimensional cylindrical sheet and a three-dimensional one, which drives this compound to be an anisotropic but three-dimensional metal \cite{singh2014multiband}. This is also consistent with the results of  upper critical field measurements where a moderate anisotropy of $\mu_0H_{c2}$ was observed and the coherence lengths along all three axes are much larger than the interchain distance \cite{jiao2015multiband}. High field measurements uncover a quasi-linear magnetoresistance without any sign of saturation up to about 50~T, as well as a violation of Kohler's rule, indicating the existence of charge density wave (CDW) fluctuations in this compound \cite{xu2015quasi}, which has also been suggested from scanning tunneling spectroscopy (STS) experiments \cite{fan2015scanning}. Meanwhile, there have also been various studies to characterize the superconducting order parameter, which could give the crucial information about the pairing mechanism. However, no firm conclusions have been reached on the nature of the gap structure. Evidence for nodal superconductivity came from  thermal conductivity measurements, where in zero field there is a significant residual value of $\kappa(T)/T$ at zero temperature. Furthermore, there is a rapid increase of $\kappa(H)/T$ in an applied magnetic field, which is very similar to the behavior of $d$-wave superconductors \cite{pan2014observation}. This scenario is supported by electronic specific heat results, which show power law behavior at low temperatures and a non-linear field dependence of the Sommerfeld coefficient $\gamma(H)$ \cite{jiao2015multiband}. On the other hand, different results were obtained from STS measurements, where a BCS-like gap structure was reported by one group, whereas another report gave an indication of a highly anisotropic gap structure with gap minima or  nodes \cite{Zeng2015anisotropic,fan2015scanning}. Therefore, due to the discrepancies between different measurements of the order parameter of Ta$_4$Pd$_3$Te$_{16}$, further measurements at lower temperatures which are sensitive to low energy excitations are badly needed.

\begin{figure}[t]
\begin{center}
\includegraphics[width=0.5\columnwidth]{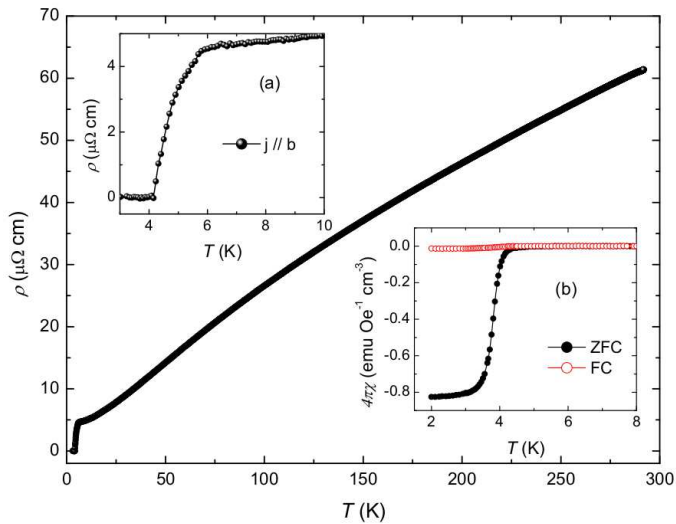}
\end{center}
\caption{The temperature dependence of the electrical resistivity $\rho(T)$ of Ta$_4$Pd$_3$Te$_{16}$ from 292K down to around 3K. The insets show the behavior of (a) $\rho(T)$ and (b) the magnetic susceptibility 4$\pi\chi(T)$ at low temperatures, where superconducting transitions are clearly observed.}
\label{Smaplecharacterize}
\end{figure}

Here, we report  the temperature dependence of the London penetration depth $\Delta\lambda(T)$ of single crystals of Ta$_4$Pd$_3$Te$_{16}$. A combined analysis of our $\Delta\lambda(T)$ measurements, the derived superfluid density $\rho(T)$ and the previous specific heat results show consistent evidence for multi-band superconductivity  in Ta$_4$Pd$_3$Te$_{16}$ with nodes in at least one of the gaps.

\section{Methods}

Single crystal samples were synthesized by a self-flux method \cite{jiao2014superconductivity}, and characterized using both electrical resistivity and magnetic susceptibility measurements. The temperature dependence of the resistivity was measured using the standard four-probe method from room temperature down to about 3~K, while the magnetization measurements were performed by utilizing a SQUID magnetometer (MPMS-5T) from 8~K to about 2~K with both field-cooling (FC) and zero-field-cooling (ZFC) under a small applied magnetic field of 10~Oe.  

By utilizing a tunnel-diode-oscillator (TDO) based technique \cite{van1975tunnel}, precise measurements of London penetration depth $\Delta\lambda(T)$ were carried out in a ${^3}$He cryostat down to 0.45~K, and in a dilution refrigerator with a base temperature of about 0.06~K. Due to the flat needle like shape of the crystals, samples were cut into typical sizes of $(250-450)\times(250-450)\times(50-150)~\mu m^3$ with the plane being parallel to the chain direction. The sample was mounted on a sapphire rod so as to be inserted into the coil without any contact. The operating frequency of the TDO was about 7~MHz in the ${^3}$He system and 9~MHz in the dilution refrigerator, with a noise level as low as 0.1~Hz, by steadily controlling the temperature of the coil and electrical circuit independently. The sample experienced a very small ac field induced by the coil  of about 20~mOe along the $c^*$ direction, which is much smaller than the lower critical field $H_{c1}$, ensuring that the sample was always in the Meissner state during the measurements. As a result, the measured frequency shift $\Delta f(T)$ can be considered to be proportional to the change of the London penetration depth in the $a^*b$ plane with $\Delta\lambda(T)$=G$\Delta f(T)$, where the calibration constant $G$  is solely dependent on the sample and coil geometry \cite{prozorov2000meissner}.

\begin{figure}[t]
\begin{center}
  \includegraphics[width=0.5\columnwidth]{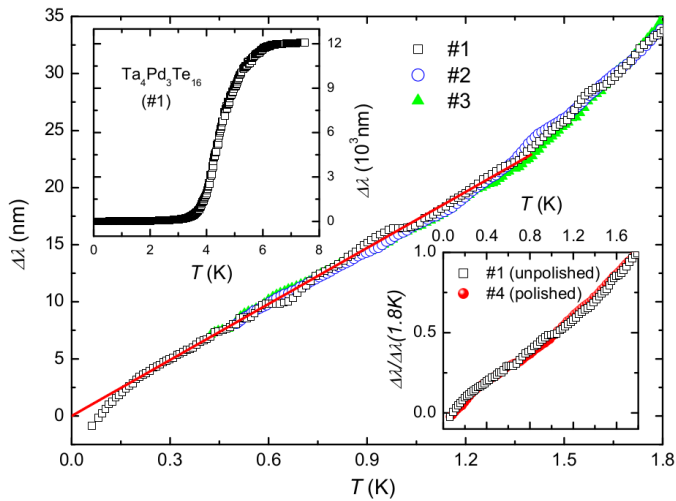}
\end{center}
	\caption{Low temperature behavior of the London penetration depth $\Delta\lambda(T)$ for three  single crystals of Ta$_4$Pd$_3$Te$_{16}$. The solid red line shows the linear decrease of $\Delta\lambda(T)$ below around 1.4~K. The upper inset shows the temperature dependence of $\Delta\lambda(T)$ up to above $T_c$. The lower inset shows a comparison between the low temperature behavior of an unpolished sample (\#1) and a sample where the surface was polished (\#4), normalized by the respective values at 1.8~K.}
   \label{deltalambda}
\end{figure}

\section{Results}

The electrical resistivity [$\rho(T)$] and magnetic susceptibility are displayed in Fig.~\ref{Smaplecharacterize}. Metallic behavior of $\rho(T)$ is shown in the normal state, with a residual resistivity of 4.5 $\mu\Omega$ cm, just before entering the superconducting state, with zero resistivity being reached at 4.1~K. This gives rise to a large mean free path of 281~nm following the method in Ref.\onlinecite{orlando1979critical}, by using a coherence length of $\xi_0$=10~nm and  Sommerfeld coefficient $\gamma_n$=51.2~mJ mol$^{-1}$ K$^{-2}$ \cite{jiao2015multiband}. This calculated mean free path is much larger than the coherence length. A superconducting transition is also observed in the magnetic susceptibility measurements, with a midpoint of the transition at around 3.8~K. These results indicate that Ta$_4$Pd$_3$Te$_{16}$ is a superconductor in the clean limit.

As shown in Fig.~\ref{deltalambda}, the temperature dependence of the London penetration depth shift $\Delta\lambda(T)$ for various single crystals was measured from 8~K down to 0.06~K, which exhibits similar reproducible behavior. A clear superconducting transition is observed in  $\Delta\lambda(T)$, with $T_c=3.9$~K determined from the endpoint of the transition, which was used in the subsequent calculations. We note that in both the  resistivity and TDO based measurements, there is an onset of superconductivity at higher temperatures of around 6~K, whereas the endpoints of the transitions are close to the $T_c$ values in the specific heat and dc magnetic susceptibility. Since the resistivity and frequency shift are much more sensitive to non-bulk superconductivity, these results suggest that there is the onset of filamentary superconductivity with a small volume fraction at higher temperatures, above the sharp onset of  bulk superconductivity in the rest of the sample. The main panel displays the enlargement of the low temperature behavior of $\Delta\lambda(T)$, where a quasilinear temperature dependence is observed from 1.4~K down to about 0.2~K, below which a small downturn in  $\Delta\lambda(T)$ is observed. The temperature dependence of $\Delta\lambda(T)$ at $T\ll T_c$ is usually related to the low energy excitations, which is determined by the superconducting gap structure. In the case of nodeless weakly coupled s-wave superconductors, $\Delta\lambda(T)$ regularly exhibits an exponential decrease below around $T_c$/3, due to the absence of low energy excitations. Whereas in other materials, such as the cuprate and heavy fermion superconductors where there are often nodes in the gap, a power law temperature dependence of $\Delta\lambda(T)$ is usually observed and in particular, when $\Delta\lambda(T)\sim$T at low temperatures, there is a strong indication of line nodes \cite{hardy1993precision,chia2003nonlocality}. Therefore, the clear observation of linear behavior at low temperatures in the penetration depth measurements is evidence against  fully-gapped superconductivity in Ta$_4$Pd$_3$Te$_{16}$, and suggests that the superconducting gap has line nodes. We note that the small downturn below 0.2~K is consistently observed in different samples, including when the sample is polished to avoid any extrinsic effect from the as-grown surface (lower inset of Fig.~\ref{deltalambda}). The overall size of the downturn of the signal is only around 0.04\% of the overall frequency shift from above $T_c$ and therefore this may correspond to the presence of a very small  impurity phase, which is not likely to significantly affect the measurements at higher temperatures.

\begin{figure}[t]
\begin{center}
  \includegraphics[width=0.5\columnwidth]{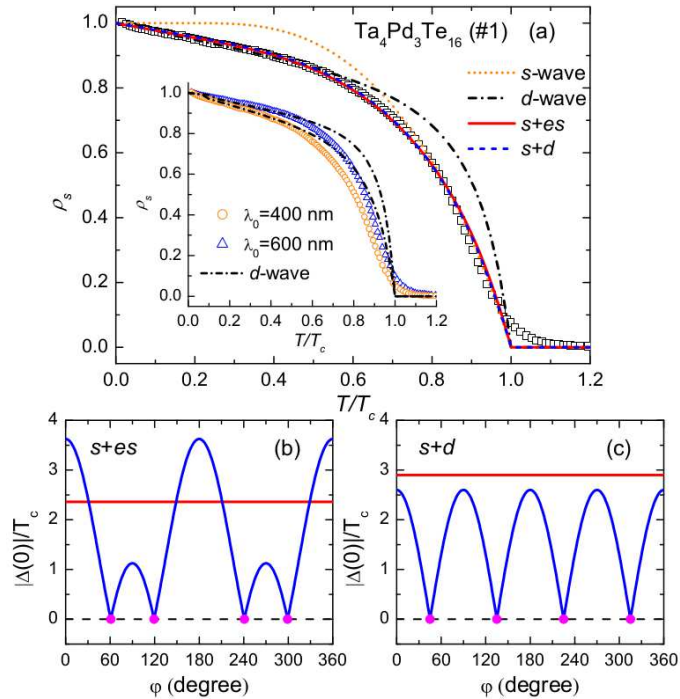}
\end{center}
	\caption{(a) The temperature dependence of the normalized superfluid density $\rho_s(T)$ for sample $\#1$ of Ta$_4$Pd$_3$Te$_{16}$, using $\lambda(0)~=~492$~nm. The lines show the results from fitting various models. The inset shows $\rho_s(T)$ for two different values of $\lambda(0)$, where dashed lines show fits to a $d$-wave model. The bottom panels show the angular dependence of the amplitudes of the two gaps at zero temperature from the fitted (b) $s+es$ and (c) $s+d$ models, where the solid dots represent the nodal directions.}
   \label{superfluid}
\end{figure}

To get further insight into the superconducting pairing symmetry of Ta$_4$Pd$_3$Te$_{16}$, the normalized superfluid density $\rho_s(T)$, calculated using $\rho_s(T)=\lambda^2(0)/\lambda^2(T)$ with $\lambda(T)=\lambda(0)+\Delta\lambda(T)$, is plotted in Fig.~\ref{superfluid}(a). The value of $\lambda(0)$ was derived from solving $\mu_0H_{c1}~=~(\Phi_0/4\pi\lambda^2)[{\rm ln}(\lambda/\xi)+0.5]$, where $\Phi_0$ is the magnetic flux quantum. By using the parameters  $\mu_0$$H_{c2}(0)$=3.3~T and $\mu_0$$H_{c1}(0)$=2.99~mT \cite{jiao2015multiband}, the value of $\lambda(0)$=492~nm was estimated. The inset of Fig.~\ref{superfluid}(a) also shows $\rho_s(T)$ for a change in  $\lambda(0)$ of $\pm\sim20\%$ and the data show similar behavior. The temperature dependence of $\rho_s(T)$ solely depends on its Fermi surface and superconducting gap structure. For a given gap function $\Delta_k$, the normalized superfluid density $\rho_s(T)$ is calculated using:
\begin{equation}
\rho_{\rm s}(T) = 1 + 2 \left\langle\int_{\Delta_k}^{\infty}\frac{E{\rm d}E}{\sqrt{E^2-\Delta_k^2}}\frac{\partial f}{\partial E}\right\rangle_{\rm FS},
\label{equation1}
\end{equation}
\noindent where $f(E, T)=[1+{\rm exp}(E/T)]^{-1}$ is the Fermi distribution function with the Boltzmann constant defined as $k_B = 1$, and $\left\langle\ldots\right\rangle_{\rm FS}$ denotes the integration over the Fermi surface. The superconducting gap function is given by $\Delta_k(T)=\Delta(T)g_k(\phi)$. Here, $g_k(\phi)$ is a dimensionless function that determines the angular dependence of the gap and $\phi$ is the azimuthal angle. $\Delta(T)$ describes the temperature dependence of the gap, which is approximated by:
\begin{equation}
\Delta(T)=\Delta(0){\rm tanh}\left\{1.82\left[1.018\left(T_c/T-1\right)\right]^{0.51}\right\},
\label{equation2}
\end{equation}
\noindent where the parameter which characterizes the zero temperature gap magnitude $\Delta(0)$ is the only adjustable parameter \cite{carrington2003magnetic}.

Several forms of $g_k(\phi)$ were used to fit the experimental data, as displayed in Fig.~\ref{superfluid}(a). Firstly, a single band $s$-wave model with $g_k(\phi)=1$ fails to describe the data at low temperatures, although the high temperature part above 0.7$T_c$ is  fitted quite well. When the superconducting gap is nodeless, $\rho_s(T)$ is always expected to flatten at $T\ll T_c$, but no such saturation is observed in the experimental data and $\rho_s(T)$ continues to increase with decreasing temperature.  The observation of a linear temperature dependence of $\Delta\lambda(T)$ and $\rho_s(T)$ provides evidence for the existence of line nodes in the superconducting energy gap, as in the case of the $d$-wave superconductivity of the cuprates. Consequently, a $d$-wave model with $g_k(\phi)={\rm cos}2\phi$ was fitted, which successfully describes $\rho_s(T)$ at low temperatures but significantly deviates in the high temperature region, implying that a single band $d$-wave model is also not a reasonable scenario. Although the derived $\rho_s(T)$ data will be affected by uncertainties in both the calibration constant $G$ and $\lambda(0)$, as displayed in the inset, upon changing $\lambda(0)$ by $\pm\sim20\%$, a $d$-wave model is still unable to account for the data.

\begin{figure}[t]
\begin{center}
  \includegraphics[width=0.5\columnwidth]{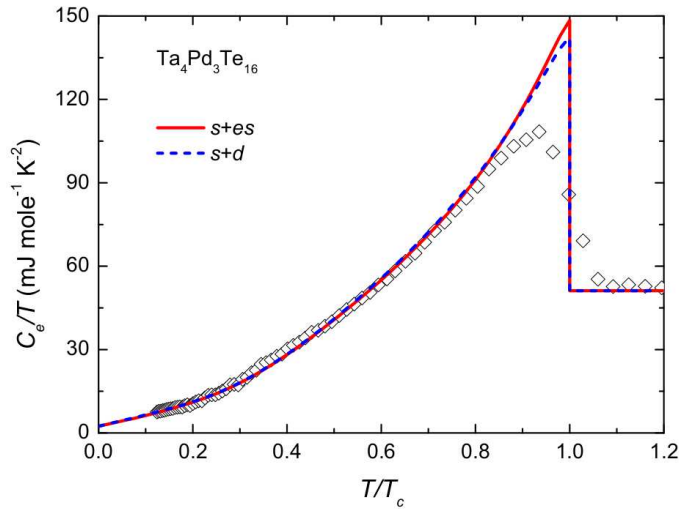}
\end{center}
	\caption{Electronic specific heat of a single crystal of Ta$_4$Pd$_3$Te$_{16}$ from Ref.~\onlinecite{jiao2015multiband}. The lines show the fits to the $s+es$ and $s+d$ models, which overlap well with the experimental data.}
   \label{specificheat}
\end{figure}

Due to the presence of multiple Fermi surface sheets from theoretical calculations \cite{singh2014multiband}, we also fitted various two-band models. In a phenomenological two-band model, the total superfluid density $\rho_s(T)$ can be obtained from a linear combination of different two components:
\begin{equation}
\rho_{\rm s}(T) = \alpha\rho{_{\rm s}^1}(\Delta_k^1, T) + (1-\alpha)\rho{_{\rm s}^2}(\Delta_k^2, T),
\label{equation3}
\end{equation}

\noindent where $\Delta_k^i (i=1, 2)$ represent the superconducting gap functions of the two components and $\alpha$ is the relative weight for $\rho_s^1$. Since a linear temperature dependence of $\rho_s(T)$ is clearly observed at $T\ll T_c$, which can not be reproduced by a calculation with two isotropic nodeless gaps, both $s+es$  and $s+d$ models were fitted to the experimental data. In the $s+es$  model there is one isotropic gap and an anisotropic $s$-wave ($es$)  gap, with a gap angular dependence given by $g_k(\phi)=1+r{\rm cos}2\phi$, where $r$ characterizes the gap anisotropy \cite{jiao2015multiband,Zeng2015anisotropic,Lin2011}. It can be seen that the energy gap is always nodeless for $r<1$, while for $r>1$,  it goes to zero along $\phi$=$0.5$arccos($-1/r$), where the accidental nodes are located. We note that a single band anisotropic $s$-wave model was previously reported to be unable to account for the specific heat data \cite{jiao2015multiband}. On the other hand the  $s+d$ model has an isotropic  $s$-wave gap as well as  a $d$-wave gap, and this model was also applied in the analysis of STS and specific heat data, where the nodal $d$-wave component was one possibility for explaining the deviation from isotropic, fully-gapped  behavior \cite{jiao2015multiband,Zeng2015anisotropic}. As shown in Fig.~\ref{superfluid}(a), both  models can well fit the experimental data across the whole temperature range. For the $s+es$  model, the fitting parameters are $\Delta_s(0)=2.36T_c$, $\Delta_{es}(0)=1.25T_c$ and $r=1.9$, with a weighting of $\alpha$=0.3 for the isotropic $s$-wave component. For the $s+d$ model, the fitted values are $\Delta_s(0)=2.9T_c$, $\Delta_d(0)=2.6T_c$ and $\alpha$=0.6. The angular dependence of the gap structures for the fitted models are displayed in Fig.~\ref{superfluid} (b) and (c), where the value of $r=1.9$ indicates that for the $s+es$  model, there are nodes in the anisotropic gap. It should be noted that the linear temperature dependence of $\rho_s(T)$ at low temperatures can not be reproduced by an $s+es$ model with $r<$1, where both superconducting gaps are fully open. In that case, due to a nodeless superconducting gap structure, $\Delta\lambda(T)$ and the derived $\rho_s(T)$ always become flat below a certain temperature depending on the value of the gap magnitude, whereas no such saturation is observed in the measurements, even down to 0.06~K ($\sim0.015T_c$).

In addition, the  $s+es$ and $s+d$  models were also fitted to the previously reported electronic specific heat data from Ref.~\onlinecite{jiao2015multiband}, as displayed in Fig.~\ref{specificheat}. The entropy $S$ in the superconducting state can be expressed as \cite{Bouquet2001}:
\begin{equation}
S = -\frac{3\gamma}{\pi^3}\int_{0}^{2\pi}\int_{0}^{\infty}[f{\rm ln}f+(1-f){\rm ln}(1-f)]\rm d\rm{\varepsilon}\rm d\rm{\phi},
\label{equation4}
\end{equation}
\noindent where $\varepsilon=\sqrt{E^2-\Delta_k^2(T)}$ and $\Delta_k(T)$ follows the same expression used in the superfluid density analysis. In the superconducting state, the superconducting electronic specific heat is derived from $C_e=T{\rm d}S/{\rm d}T$. Both the $s+es$ and $s+d$  models can well describe the data, taking into account a residual contribution to $C_e/T$  of  $\gamma_0$=2.4~mJ~mol$^{-1}$ K$^{-2}$. The fitted parameters are $\Delta_s(0)=2.1T_c$, $\Delta_{es}(0)=1.16T_c$, $\alpha$=0.53 and r=2.0 for the $s+es$  model and $\Delta_s(0)=2.45T_c$, $\Delta_d(0)=2.05T_c$, and $\alpha$=0.48 for the $s+d$ model. Since the specific heat is sensitive to excitations along all directions but $\rho_s$(T) only probes directions perpendicular to the applied field, this may account for the small differences between the fitting parameters from the two techniques.

Therefore, both two-band $s+es$ and $s+d$ models can reasonably account for the penetration depth and specific heat data, so it is difficult to distinguish between these two scenarios on the basis of our measurements. The primary result of our present study is the linear behavior of $\Delta\lambda$(T) at low temperatures, which strongly suggests the existence of line nodes in the gap structure. However, in the case of the $s+d$ model, the $s$-wave and $d$-wave instabilities will generally be expected to have different transition temperatures and therefore a split superconducting transition would be anticipated \cite{Sig2000}. It has also been noted that the nodal superconducting gap structure is expected to be more robust in $d$-wave superconductors than in the extended $s$-wave cases, where the accidental nodes can be easily lifted by the effect such as disorder \cite{reid2012d}. Therefore, it maybe helpful to study samples where random defects are introduced, for example by electron irradiation. On the other hand,  theoretical calculations indicate that there are complex anisotropic Fermi sheets in Ta$_4$Pd$_3$Te$_{16}$, and suggest the system is far away from a magnetic instability \cite{singh2014multiband}. In that sense, the complex Fermi surface maybe play an important role in leading to the anisotropic interactions which give rise to nodes.

\section{Conclusion}

To summarize, we have precisely measured the temperature dependence of the change of the London penetration depth $\Delta\lambda(T)$ for the newly discovered layered superconductor Ta$_4$Pd$_3$Te$_{16}$ using a TDO method. Linear behavior of $\Delta\lambda(T)$ is clearly observed at low temperatures, as well as in the corresponding superfluid density $\rho_s(T)$, which can be successfully described in terms of either a phenomenological two-band $s+es$ or $s+d$ model, in line with our reanalysis of the previous specific heat results. Our findings show a distinct discrepancy from the behavior of fully-gapped superconductors and provide strong evidence for nodal superconductivity in Ta$_4$Pd$_3$Te$_{16}$.
\section{Acknowledgments}

We thank X.~Lu, C. Cao and D. F. Agterberg for helpful discussions and suggestions. This work was supported by the National Key R\&D Program of China (No.~2016YFA0300202,No.~2017YFA0303100,), the National Natural Science Foundation of China (No. 11474251), and the Science Challenge Project of China (No. TZ2016004).

\end{document}